\documentclass[a4paper,11pt]{article}
\pdfoutput=1 

\usepackage{jinstpub} 

\title{\boldmath A large facility for photosensors test at cryogenic temperature}

\author[a,c]{Z.~Balmforth,}
\author[b]{A.~Basco,}
\author[b]{A.~Boiano,}
\author[b,1]{N.~Canci,\note{Corresponding author.}}
\author[b]{R.~de~Asmundis,}
\author[c,b]{F.~Di~Capua,}
\author[c,b]{G.~Fiorillo,}
\author[b]{G.~Grauso,}
\author[c, b]{G.~Matteucci,}
\author[b]{A.~Pandalone,}
\author[d,c]{E.~Sandford,}
\author[c,b]{Y.~Suvorov,}
\author[b]{G.~Tortone,}
\author[b]{A.~Vanzanella,}

\affiliation[a]{Department of Physics, Royal Holloway University of London,\\Egham TW20 0EX, UK}
\affiliation[b]{INFN - Napoli,\\Napoli 80126, Italy}
\affiliation[c]{Physics Department, Universit\`a degli Studi di Napoli Federico II,\\Napoli 80126, Italy}
\affiliation[d]{Department of Physics and Astronomy, University of Manchester, \\Manchester M13 9PL, UK}

\emailAdd{nicola.canci@infn.it}

\abstract{Current generation of detectors using noble gases in liquid phase for direct dark matter search and neutrino physics need large area photosensors.
Silicon based photo-detectors are innovative light collecting devices and represent a successful technology in these research fields. 
The DarkSide collaboration started a dedicated development and customization of SiPM technology for its specific needs resulting in the design, production and assembly of large surface modules of 20$\times$20~cm$^{2}$ named Photo Detection Unit for the DarkSide-20k experiment. 
Production of a large number of such devices, as needed to cover about 20~m$^{2}$ of active surface inside the DarkSide-20k detector, requires a robust testing and validation process. 
In order to match this requirement a dedicated test facility for the photosensor test was designed and commissioned at INFN-Naples laboratory. The first commissioning test was successfully performed in 2021. Since then a number of testing campaigns were performed. 
Detailed description of the facility is reported as well as results of some tests.}

\keywords{Photosensors; Silicon Photomultipliers; Cryogenics; Liquid argon; Noble liquid detectors (scintillation, ionization, double-phase)}

\arxivnumber{4634179} 

\collaboration[c]{on behalf of the DarkSide Collaboration}

\proceeding{LIDINE2022\\
  September 21-23, 2022\\
  University of Warsaw Library}

\begin{document}
\maketitle
\flushbottom


\section{Introduction}
\label{sec:intro}

The use of liquified noble gases (neon, argon and xenon) as sensitive medium offers one of the most promising detection techniques to investigate the fields of direct dark matter search and neutrino physics. 
Due to their features such as high scintillation photon yield, powerful background rejection through efficient event discrimination methods, ease of purification, noble liquids represent an ideal medium for these experimental searches~\cite{doke, kubota1, kubota2}.
Scintillation light from noble liquids (converted by appropriate wavelenghtshifting material in the case of argon) can be detected by photosensors operating at low temperature. 
Silicon photosensors, such as Silicon PhotoMultipliers (SiPMs), are well suited to this purpose~\cite{sipm_ds1, sipm_ds2, sipm_lar1, sipm_lar2, sipm_lxe1, sipm_lxe2, sipm_lxe3}.\\
SiPM devices represent one of the key elements of the DarkSide-20k (DS-20k) experiment, to be installed in the Hall C of the INFN-LNGS laboratory for the direct dark matter search using radiopure (low $^{39}$Ar content) argon from underground sources~\cite{acosta, ds50_uar_first, ds50_uar_532}.  
The dual phase Time Projection Chamber (TPC) with $\sim$50~t radiopure argon mass represents the core of the apparatus and serves as both target and detector to observe dark matter particles scattering from argon nuclei~\cite{ds_tdr}. 
It is instrumented with large area SiPMs arrays, called Photo Detection Units (PDUs), to detect both the prompt argon scintillation photons produced in the liquid (S1) and the delayed light from electroluminescence produced by the ionization electrons drifted to and extracted in the gas region (S2). 
To this purpose a large number of PDUs able to cover about 20~m$^{2}$ of active surface inside the DarkSide-20k detector, will be constructed. Their production will take place in Nuova Officina Assergi (NOA) a dedicated assembly facility for silicon photosensors located in the Hall di Montaggio at the external laboratories of INFN-LNGS.
All the PDUs produced at NOA will undergo a full characterization test in liquid nitrogen (LN) before installation in the DS-20k TPC. A dedicated test facility has been built at INFN-Naples laboratory to this specific purpose, where commissioning and several tests have already been successfully performed. The system also allows for the characterization in liquid argon and can host any other type of large area photosensor as required by the present generation of experiments based on noble liquids technology.
%
%
%
%
%
%
%
%

\section{Facility description}
\label{sec:ptf_description}

\begin{figure}[!ht]
\begin{center}
\includegraphics*[width=9.3cm,angle=0]{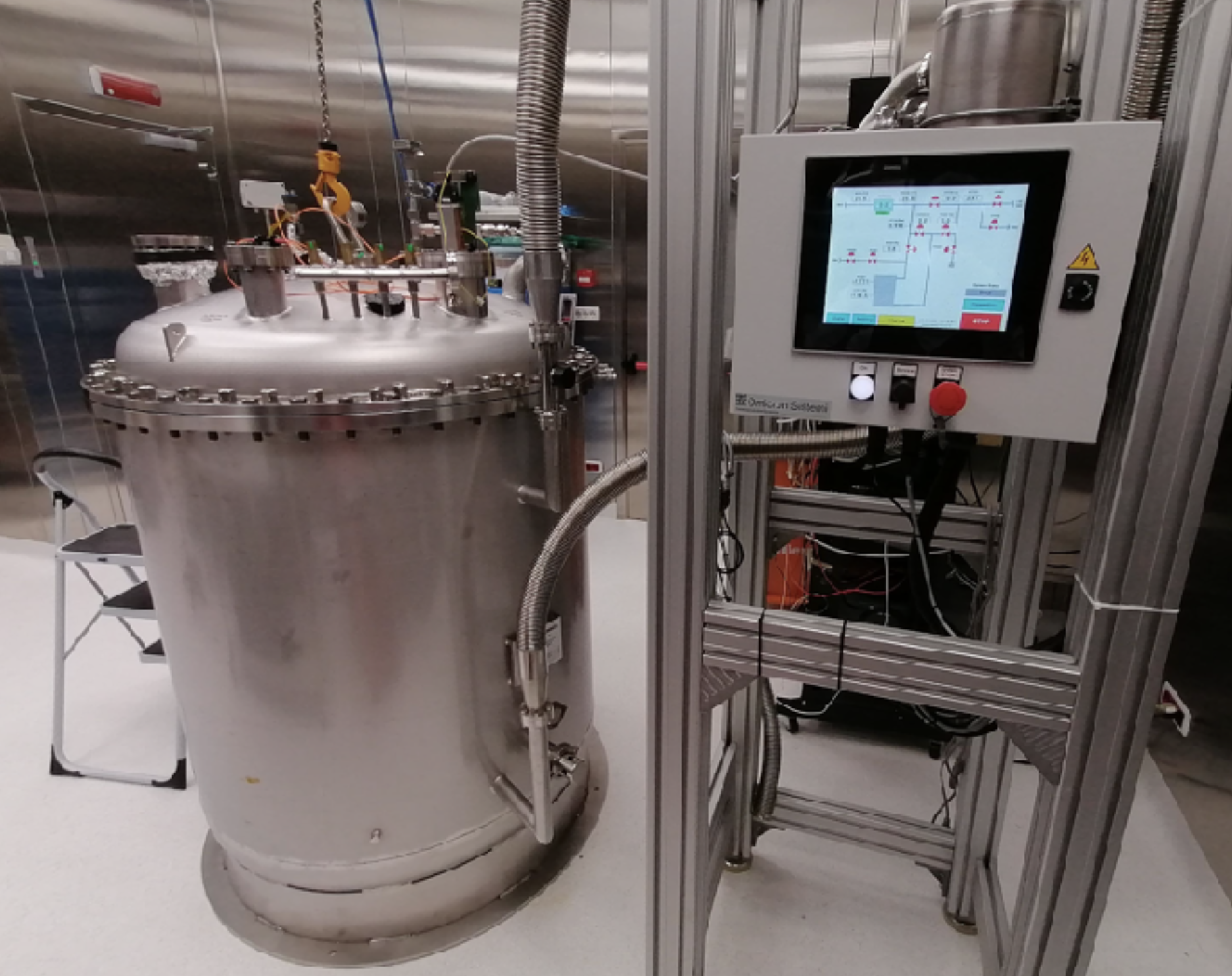}
\includegraphics*[width=5.5cm,angle=0]{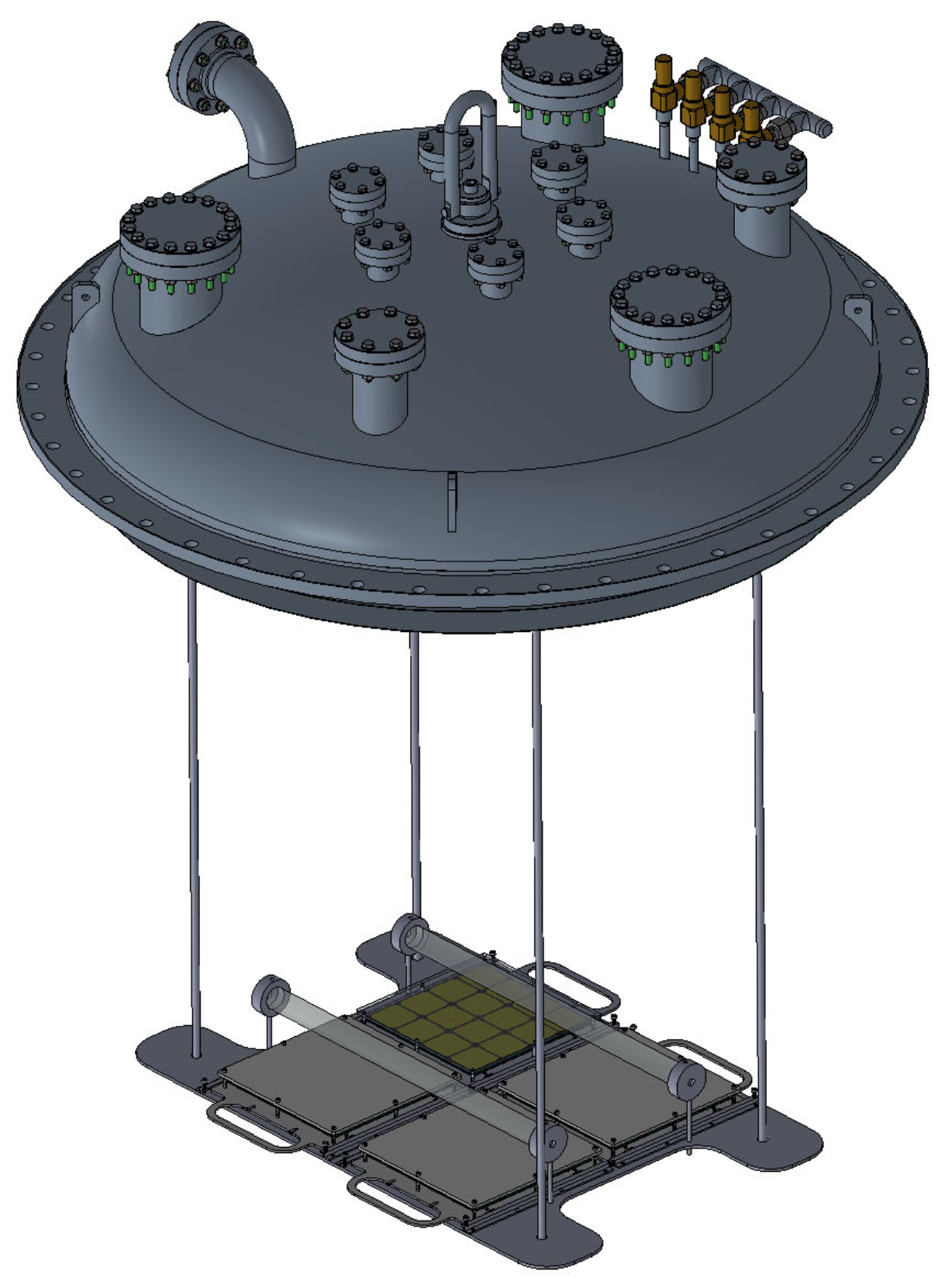}
 \caption{Picture of the Photosensor Test Facility located in the clean room of the Naples cryogenic laboratory [Left]. Conceptual design of the mechanical structure with the PDUs and the illuminating system [Right].}
\label{fig:ptf_facility}
\end{center}
\end{figure}

The Photosensor Test Facility (PTF) is hosted in the cleanroom of the INFN-Naples cryogenic laboratory.
The system is a custom made cryostat with a maximum allowed absolute pressure of 3~bar. It is  composed of a 900~l double wall vacuum insulated vessel (1.15~m diameter - 1.57~m height) coupled with a single wall domed flange and it is equipped with PT100 thermistors, a level meter, pressure transducers, analog pressure indicators and a set of pressure relief valves for safety.
The top flange is provided with different CF flanges for electrical and optical feedthroughs.
Two ports connected to the double wall vessel are used respectively for the liquid nitrogen (or argon) inlet and cold gas venting.
The inlet and outlet are connected to a vacuum insulated vessel containing proportional and pneumatic valves, temperature and pressure sensors, the so called valve box, automatically operating the filling and draining processes~(fig.~\ref{fig:ptf_facility}~[left]).\\
The facility cryostat is connected to an external 3000~l liquid nitrogen storage tank with an operating pressure of 2.7~bar.
The system controls, namely proportional and automatic valves and the operating parameters (pressure, temperatures, nitrogen level, etc.) are controlled by a PLC-based control unit and monitored by a National Instruments PXI computer and a LabView-based program.
The pressure in the cryostat after filling is maintained at a constant value of 1100~mbar.\\
A dedicated mechanical internal structure designed to host the photosensors is hanged to the domed flange through four PEEK insulating adapters for reducing the thermal conduction.\\
For the DarkSide application, the mechanical structure is made of three adjustable horizontal planes connected with four threaded rods hosting four PDUs, each positioned face-up, for a total of 12 PDUs.
The PDUs are previously mounted on customized support plates allowing for insertion and removal from the mechanical structure without any dismounting~(fig.~\ref{fig:ptf_facility}~[right]).\\
Two different light distribution systems are available, based on acrylic rods and PTFE diffusers respectively. They are placed above each of the photosensor plates. They are connected to optical fibers transmitting the light emitted by a Hamamatsu PLP Laser at 407~nm. The illuminating system can be adjusted in distance with respect to the PDUs plane to optimize the uniformity of the light distribution.\\
The electronics is placed in a rack positioned close to the cryostat. The bias voltages are provided by a CAEN mainframe hosting the low and high voltage power supplies, while signals from the photosensors are acquired by 64 channels CAEN VX2740 waveform digitizers installed in a VME crate.
Three computing servers provide for data acquisition, data storage and data analysis. 
A dedicated software, based on the MIDAS framework~\cite{midas}, gives fully integrated controls for power supplies, DAQ, e-logs, data summaries, slow control history,  cryogenic parameters and online monitoring and allows for remote and automated data acquisition runs.\\
The filling process starts with a cooling phase of the transfer lines, valve box and cryostat obtained by flushing cold 
nitrogen gas through the system decreasing the temperature while LN comes out from the storage tank and gradually cools down the piping.
The pressure in the cryostat is kept stable at the set point (1100~mbar) by acting on the apertures of the proportional valves during the filling phase where a filling rate of 1~cm/min has been achieved in the test.
Once the desired level in the cryostat is reached the filling process ends. Due to the heat exchange with the external ambient and the heat load inside the cryostat the LN undergoes continuous evaporation with the measured rate of 0.2~cm/h. After the LN level reaches a defined low level set point the automated refill procedure starts and, after the cooldown, LN is added until the previous threshold level is reached. 
At the end of the test cycle the cryostat can be drained with a dedicated automated procedure. 
As the first step the vent line is closed and the cryostat is pressurized up to 2.7~bar to push the LN out through the inlet transfer line into a heat exchanging evaporator located in the external facilities area of the laboratory and then discharged in atmosphere in form of gas. 
A maximum draining speed of 0.2 cm/min was observed (based on the over pressure created in the cryostat). 
The draining process automatically stops when the room temperature is recorded by the temperature sensor located at the bottom of the vessel. At this point the cryostat can be opened.
%
%
%
%
%
%
%
%
%
%
%

\section{Test of Photosensors}
\label{sec:pdu_test}

\begin{figure}[!ht]
\begin{center}
\includegraphics*[width=6.1cm,angle=0]{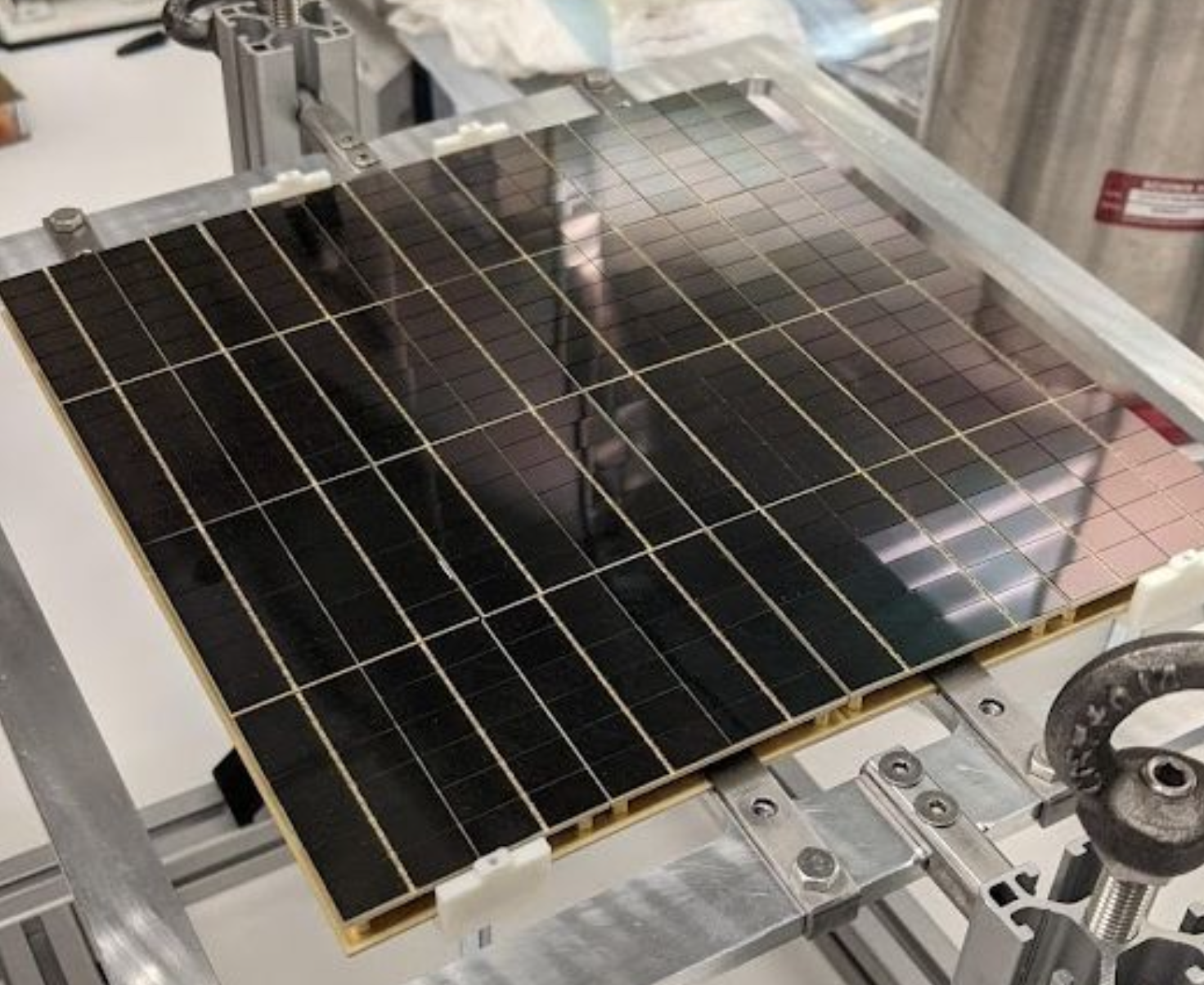}
\includegraphics*[width=5cm,angle=0]{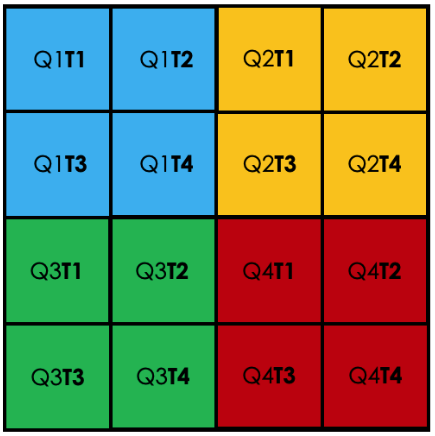}
\includegraphics*[width=3.78cm,angle=0]{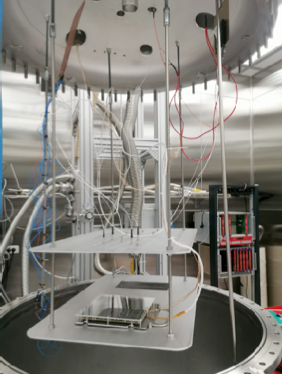}
 \caption{Picture of the Photo Detection Unit [Left]. Conceptual representation of the PDU with 4 quadrants [Center]. Picture of the PDU mounted on the mechanical structure [Right].}
\label{fig:pdu}
\end{center}
\end{figure}

As already mentioned the PTF is capable to accomodate different types of photosensors for characterization in cryogenic environment. We report here the measurements performed on the first prototype Photo Detection Unit of the DarkSide-20k detector.\\
The PDU integrates 16 SiPM tiles grouped in 4 quadrants forming an array with dimensions 20$\times$20~cm$^{2}$ and 4 readout channels~(fig.~\ref{fig:pdu}~[left]-[center]).
The PDU was produced in NOA, transported to INFN-Naples cryogenic laboratory and installed into the Photosensor Test Facility for testing~(fig.~\ref{fig:pdu}~[right]). The main goal of the test campaign was to investigate the photosensor key features and their stability on the long term period of about one month.\\
The PDU was illuminated with 4 laser diffusers from the optical plane placed at a distance of 10 cm with the light emitted by a Hamamatsu PLP-10 pulsed laser diode.
Data were acquired with both random and laser triggers. 
The recorded events were stored in MIDAS format and were analyzed and reconstructed by a dedicated software allowing for both online monitoring and offline reconstruction. For some of the offline analyses, filtering algorithms were applied to the data prior to extract the desired parameters.
%
%
%
%
%
%
%
%
%
%
%
\section{Characterization of PDUs}
\label{sec:pdu_test_2}

\begin{figure}[h!]
\begin{center}
\includegraphics*[width=7.8cm,angle=0]{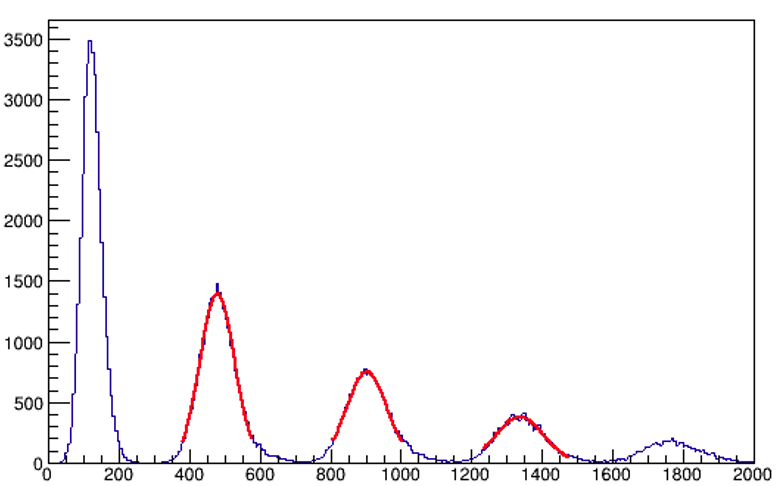}
\includegraphics*[width=7.2cm,angle=0]{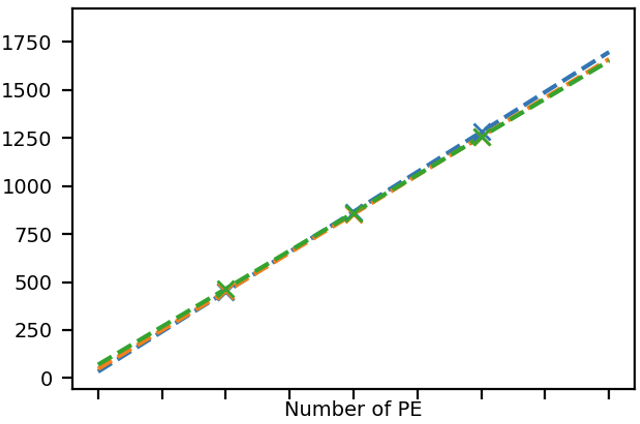}
 \caption{Single photoelectron distribution [Left] and extrapolated gain [Right] for one of the quadrants of the PDU at LN temperature.}
\label{fig:gain}
\end{center}
\end{figure}

\begin{figure}[h!]
\begin{center}
\includegraphics*[width=15cm,angle=0]{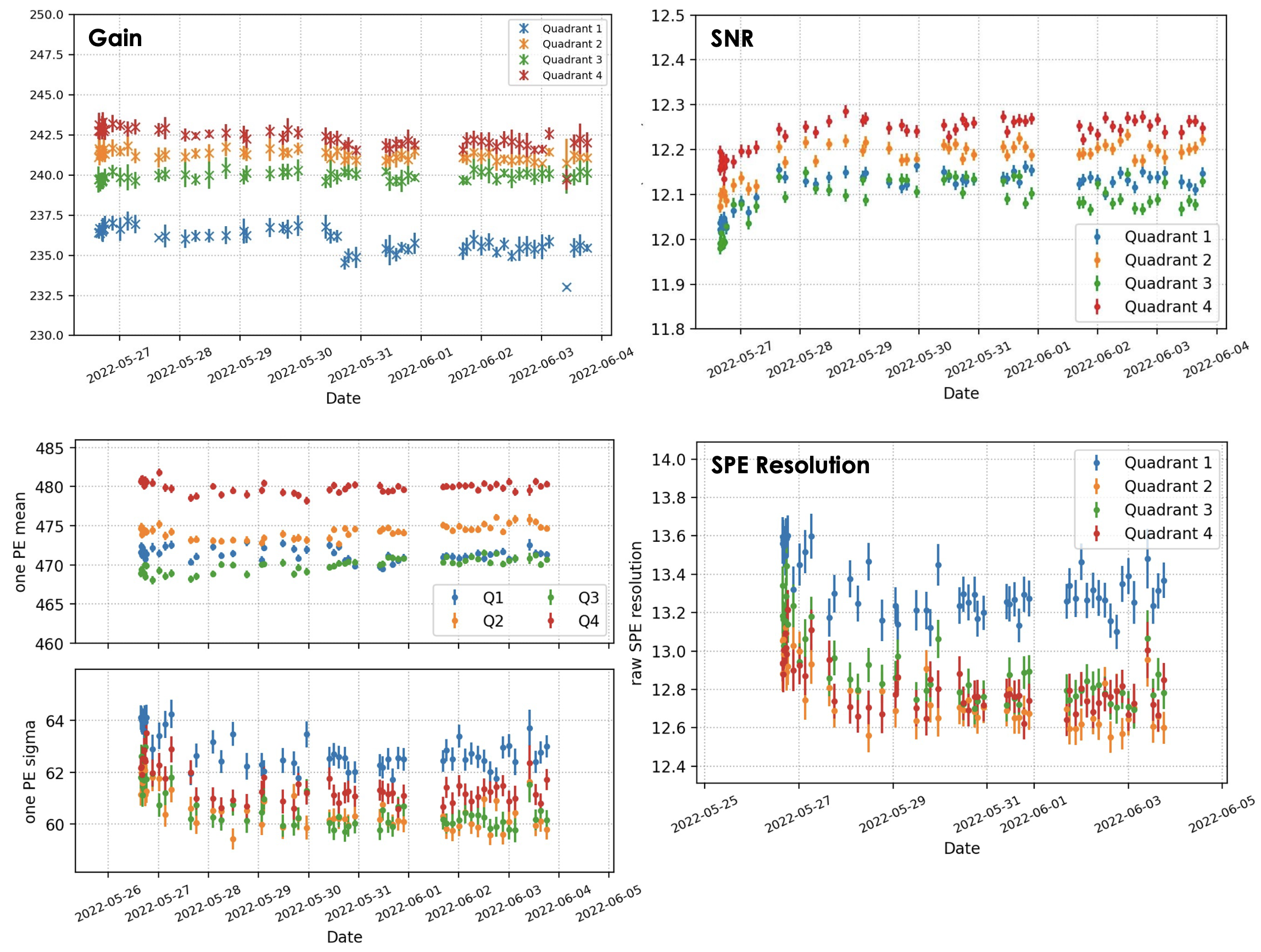}
 \caption{Behaviours of the gain [Top - Left], SNR [Top - Right] and SPE resolution [Bottom] for the 4 quadrants of the PDU as function of the time.}
\label{fig:stability_4}
\end{center}
\end{figure}

\begin{figure}[h!]
\begin{center}
\includegraphics*[width=9cm,angle=0]{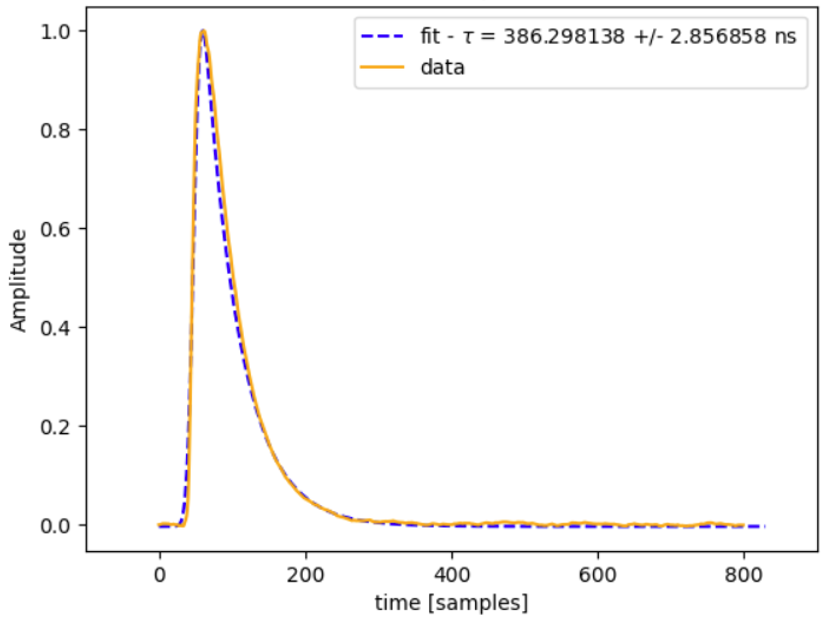}
 \caption{Average waveform with superimposed fit for one of the quadrants of the PDU.}
\label{fig:pulse_shape}
\end{center}
\end{figure}

The first step of the test consisted in the measurement of IV curves at room temperature and in liquid nitrogen. The breakdown voltages of the SiPMs were determined by a fit to the measured curves. \\
Successively, with the PDU immersed in liquid nitrogen, the following measurements were performed  for several overvoltages and at regular time intervals: amplitude and charge spectra produced with pulsed laser light, single photoelectron response (SPE), gain, single photoelectron pulse shape, signal-to-noise ratio (SNR), noise spectra, dark counts (DCR).\\
Both charge and amplitude are evaluated for each waveform in a fixed region of the acquisition window (ROI i.e. region of interest), defined to fully contain the 1 PE signal. The charge is measured by integrating the baseline subtracted waveform in the ROI.
In order to measure the gain, a gaussian model is fit to the first three peaks of the charge spectrum, shown in~fig.~\ref{fig:gain}~[left]. The mean value of each peak is plotted against the number of photoelectrons and the gain is found from the fitted slope as reported in~fig.~\ref{fig:gain}~[right].\\
The SNR is evaluated as the ratio of the average maximum amplitude of the single photoelectron waveform over the RMS of the baseline.\\
Finally, the SPE resolution is computed as the ratio of the sigma over the mean value both obtained from a gaussian fit of the first photoelectron peak in the amplitude spectrum.\\
The stability in time of the gain, SNR and SPE resolution were monitored and each of the observed parameters resulted stable within $\sim$3\% RMS over a period of several days as shown in~fig.~\ref{fig:stability_4}.\\
The pulse shape of the single photoelectron signals was studied by averaging waveforms with an amplitude compatible with 1 PE. 
A model function accounting for the discharge, quenching and recharge time of the SiPMs and the effects of the front-end electronics is then fitted to the average waveform to obtain the time decay constants as shown in~fig.~\ref{fig:pulse_shape}.
The model function uses a convolution of the functions reported below: 
$$    \text{fast}=\frac{1}{\sqrt{2\pi}\sigma}e^{-\frac{(x-x_{0})^{2}}{2\sigma^{2}}}~~~~~~\text{slow}=\frac{1}{2\tau}e^{-\frac{1}{\tau}(x-x_{0}-\frac{\sigma^2}{2\tau})}\left(1+errf\left(\frac{x-x_{0}-\frac{\sigma^2}{2\tau}}{\sqrt{2\pi}\sigma}\right)\right)$$
where $\sigma$ is the standard deviation of the gaussian fit, $x_0$ is the arbitrary $x$ offset of the peak, $\tau$ is the decay time of the exponential component, and $errf$ is the error function.\\
As a last step, dark count rates were measured by counting the number of pulses detected in a fixed  interval in the pretrigger region of the acquisition window, normalized to the total considered time interval.
DCR were estimated for all the quadrants both in laser runs and in dedicated runs with random trigger with similar results. 
%
%
%
%
%
%
%
%
%
%
%
%
%
%
%
%
%
%
%
%
%
%

%
\section{Conclusions}
\label{sec:concl}
A large Photosensor Test Facility, PTF, designed to perform tests at cryogenic temperature of large area photosensors, has been assembled, commissioned and is ready at INFN-Naples cryogenic laboratory.
Testing campaigns of large assembled arrays of SiPMs have been performed with the aim of characterizing these devices in terms of the main relevant photosensor parameters.
The dimensions and the features of the facility make it a versatile system able to host different types of photosensors to be characterized in low temperature environment with the use of liquid nitrogen or liquid argon, operating on several photosensors per cycle and allowing for weekly or even longer runs.




\acknowledgments

The authors would like to thank the Physics Department of the Universit\`a degli Studi di Napoli Federico II, the Department of Physics of the Royal Holloway University of London and the Department of Physics and Astronomy of the
University of Manchester for the agreements of cooperation between the institutions supporting this work. We also acknowledge Eng.~C.~Kendziora, Dr.~H.~Wang and Eng.~M.~Carlini for the productive discussions on the cryogenics, P.~Amaudruz, A.~Capra and B.~Smith for the work performed on the data acquisition system.



\end{document}